\documentclass[aps,pra,twocolumn,groupedaddress]{revtex4}

\usepackage{graphicx}
\usepackage{dcolumn}
\usepackage{bm}
\usepackage{amssymb}
\usepackage{subfig}
\usepackage{latexsym}
\usepackage{enumerate}
\usepackage[dvipdfmx]{color} 
\usepackage{amsmath}
\usepackage[dvipdfmx]{epsfig}
\usepackage{times}
\usepackage{caption}

\begin{document}
\widetext
\title{Blind quantum computation over a collective-noise 
channel}
\author{Yuki Takeuchi${}^{1}$, Keisuke Fujii${}^{2,3}$, Rikizo Ikuta${}^1$, Takashi Yamamoto${}^1$, and Nobuyuki Imoto${}^1$}
\affiliation{$^1$Graduate School of Engineering Science, Osaka University, Toyonaka, Osaka 560-8531, Japan\\$^2$The Hakubi Center for Advanced Research, Kyoto University, Yoshida-Ushinomiya-cho, Sakyo-ku, Kyoto 606-8501, Japan\\$^3$Graduate School of Science, Kyoto University, Kitashirakawa Oiwake-cho, Sakyo-ku, Kyoto 606-8502, Japan}

\begin{abstract}
Blind quantum computation (BQC)
allows 
a client (Alice), who only possesses relatively poor quantum devices, 
to delegate universal quantum computation to a server (Bob) in such a way that Bob cannot know Alice's inputs, algorithm, and outputs.
The quantum channel between Alice and 
Bob 
is noisy,
and the loss over the long-distance quantum communication
should also be taken into account. 
Here we propose to use decoherence-free 
subspace (DFS) to overcome the collective
noise in the quantum 
channel
for BQC, which we call DFS-BQC.
We propose three variations of DFS-BQC protocols.
One of them, a coherent-light-assisted DFS-BQC protocol,
allows Alice to faithfully send 
the signal photons with a probability proportional to a transmission rate of the quantum channel. 
In all cases, we combine 
the ideas based on 
DFS and the Broadbent-Fitzsimons-Kashefi protocol, which is one of the BQC protocols, without degrading unconditional security.
The proposed DFS-based schemes are generic and hence
can be applied to other BQC protocols
where Alice sends quantum states to Bob.
\end{abstract}

\maketitle
\begin{center}
\bf{I. INTRODUCTION}
\end{center}

A first-generation fully fledged quantum computer will eventually be realized 
by a large enterprise or a government.
It is supposed that due to its scale and/or the difficulty of maintaining it, clients who want to utilize the quantum computer will 
delegate quantum computation 
to the quantum server using poor quantum devices that are insufficient for universal quantum computation. 
In such a situation, 
the clients can employ blind quantum computation (BQC) to 
guarantee the 
unconditional
security of their inputs, algorithms, 
and outputs of quantum computations
~\cite{[C05],[AS06],[BFK09],[ABE10],[MDK10],[DKL12],[M12],[MF12],[FK12],[MF13PRA],[SKM13],[MF13PRL],[GMMR13],[MDF13],[FBSYLPJR14],[M14],[KKD14],[DF14],[HDF15],[GKW15]}.

The early BQC protocol 
proposed by 
Childs is based on the circuit model 
and assumes a client (Alice) possesses a quantum memory~\cite{[C05]}.
Broadbent, Fitzsimons, and Kashefi (BFK) 
proposed a BQC protocol 
based on the idea of measurement-based quantum computation 
(MBQC) \cite{[RBB03]},
which successfully allows Alice to be almost classical, 
only requiring a preparation of rotated qubits 
and the ability to access a quantum channel~\cite{[BFK09]}.
The BFK protocol has stimulated 
the community and has 
led to a
series of proposals of BQC 
based on the idea of MBQC
~\cite{[MDK10],[DKL12],[M12],[MF12],[FK12],[MF13PRA],[MF13PRL],[MDF13],[M14],[DF14],[HDF15],[GKW15]}.
Furthermore,  
proof-of-principle experiments 
have already been demonstrated
~\cite{[BKBFZW12],[BFKW13]}.

In 
single-server BQC protocols~\cite{[C05],[AS06],[BFK09],[ABE10],[MDK10],[DKL12],[M12],[MF12],[FK12],[MF13PRA],[SKM13],[GMMR13],[MDF13],[FBSYLPJR14],[M14],[KKD14],[DF14],[HDF15],[GKW15]}, 
Alice and a quantum server (Bob) need to execute quantum communication between them.
In such protocols,
a quantum channel between Alice and Bob is assumed to be ideal 
as long as quantum states are transmitted without loss.
This is an undesirable assumption, since an actual quantum channel has noise. Moreover, it is known that quantum communication is essential for BQC~\cite{[MK14]}.

There are several ways to fix this issue, as follows. 
First, Alice and Bob may perform entanglement distillation to share high-fidelity Bell pairs 
between them. However, in such a case, Alice has to perform quantum operations,
which are too demanding in the BQC scenario. 
Second, the parties may employ fault-tolerant topological BQC on Bob's fully fledged quantum computer to correct errors during the quantum communication~\cite{[MF12]}. 
While the threshold value $0.43\%$ 
of the error rate per gate
would be high enough for the fully fledged quantum computer on Bob's side, 
it seems to be too low to tolerate the noise introduced during the
long-distance quantum communication. 
Third, the parties may utilize double-server BQC~\cite{[MF13PRL]}, where two 
servers initially share nonmaximally entangled states due to the noise in the quantum channel, 
no quantum communication is required between Alice and two 
servers, 
and they employ entanglement distillation between two servers. 
However, in the double-server BQC protocol, 
any communication between two 
servers is prohibited. 
If two quantum servers 
communicate with each other, Alice's secrets are completely exposed to them. 
Accordingly, a complete solution of the noise problem of the quantum channel 
in BQC is still open.

In this paper, we resolve the noise problem in the quantum channel for BQC, specifically for the collective noise, which is a prototypical model of noise, as confirmed in experiments~\cite{[SGGRZ02]}, when photons are commonly used as carriers of information in quantum communication, and optical fibers are employed as quantum channels.
Decoherence-free subspace (DFS) has been known to be 
immune to such 
noise~\cite{[DG97],[ZR97],[BGLPS04],[YSOKI05],[KYKI13]}
and its validity has already been demonstrated experimentally~\cite{[MLRS03],[KBAW00],[BEGKCW04],[CZBJYZYLP06],[YNSOKI07],[PTSPKZ07],[YHOKI08],[IOTYKI11]}.

Here we propose protocols to employ DFS 
for BQC, namely, DFS-BQC protocols.
We show that parties can protect the quantum state 
sent from Alice to Bob against an arbitrary collective noise 
with few changes on Alice's side and quantum communication parts of the 
BFK protocol,  while Bob needs to perform additional operations. Since the 
BFK protocol has unconditional security against Bob's arbitrary operations, this construction substantially relaxes the proof of blindness of DFS-BQC protocols.

We propose three variations of DFS-BQC protocols.
The first protocol is an entanglement-based DFS-BQC protocol,
where Alice is required to be able to 
generate a Bell pair. However, in the BQC scenario, this requirement is too demanding for Alice.
The second one, 
a single-photon-based DFS-BQC protocol 
successfully replaces the entanglement generation process
with a single-photon source
and a postselection on Bob's side.
The third one is a coherent-light-assisted DFS-BQC protocol,
where a single photon for utilizing the DFS in the second one is replaced by a coherent-light pulse. This replacement improves the efficiency of this protocol 
under a lossy quantum channel. These protocols require only linear optics to Alice after the state preparation and do not employ single-photon interference.

This paper is organized as follows: In Sec. II, we introduce a practical noise model in the transmission channel, the procedure of the BFK protocol, and the essential properties of the BQC protocols (correctness and blindness). In Sec. III, we propose the entanglement-based DFS-BQC protocol as the first protocol. In Sec. IV, we propose the single-photon-based DFS-BQC protocol as the second protocol. In Sec. V, we propose the coherent-light-assisted DFS-BQC protocol as the third protocol. Section VI is devoted to the conclusion. In Appendices A, B, and C, we provide the details of the proof of correctness for each protocol. In Appendix D, we provide the detailed calculation of the success probability of the coherent-light-assisted DFS-BQC protocol.

\medskip
\begin{center}
\bf{II. SETUP}
\end{center}

\begin{figure}[t]
 \begin{center}
\includegraphics[width=8.6cm, clip]{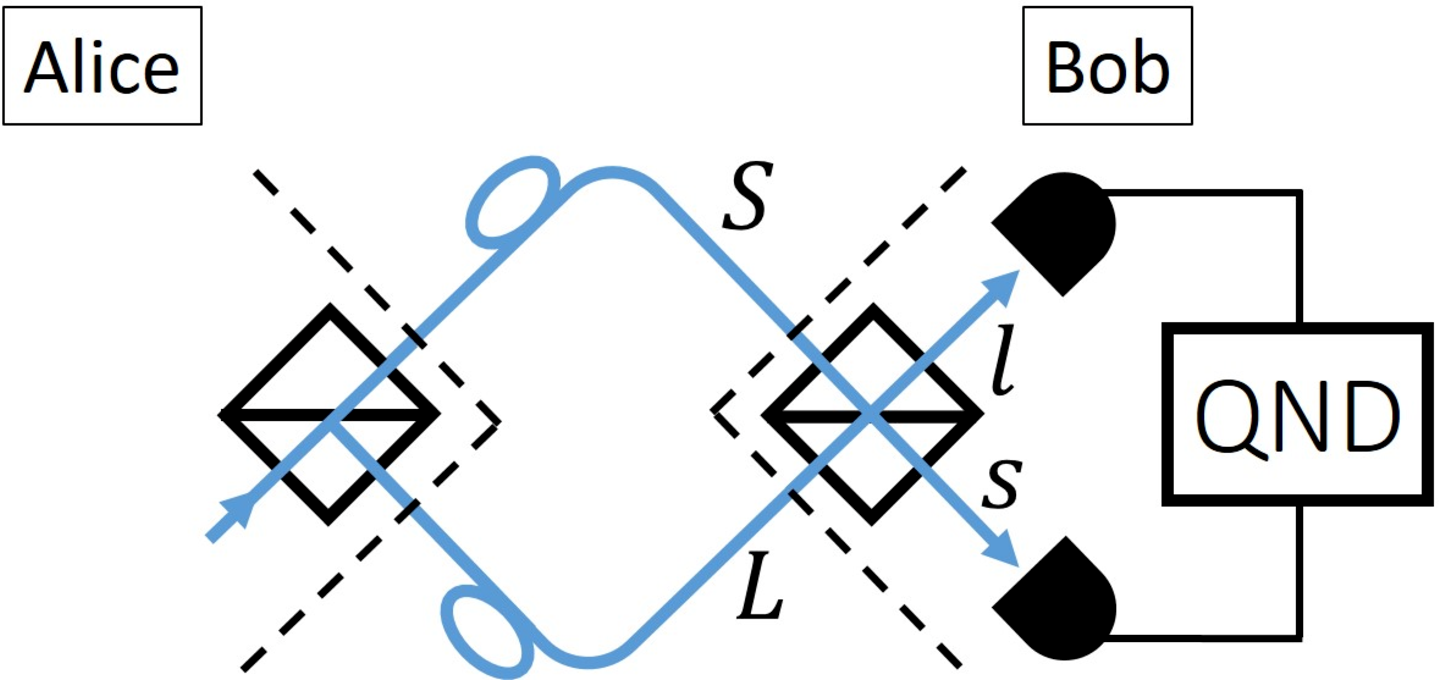}
\end{center}
\captionsetup{justification=raggedright,singlelinecheck=false}
\caption{
The 
quantum channel between Alice and Bob.
The boxes with the diagonal line indicate the polarizing 
beam splitters (PBSs).}
\label{PMZI}
\end{figure}

We employ the polarization degree of freedom of a single photon
as a qubit
$\alpha |H\rangle _m+ \beta |V\rangle _m$ ($|\alpha|^2+|\beta|^2=1$, $\alpha,\beta\in\mathbb{C}$),
where $m$ indicates the spatial mode, and $|H\rangle$ and $|V\rangle$ represent the horizontal ($H$) and vertical ($V$) polarization states of the single photon, respectively.
We may switch the notation $|H\rangle$ and $|V\rangle$ to $|0\rangle$
and $|1\rangle$, respectively,
to define the Pauli operators and the controlled-NOT (CNOT) gate.
Instead of sending such a photonic qubit directly,
Alice sends the photonic qubit through optical fibers of the transmission rate $T$ after splitting them into two spatial modes $S$ and $L$ by a polarizing beam splitter (PBS), as shown in Fig. \ref{PMZI}~\cite{[YSOKI05]}.
If the optical fibers are ideal,
the state after Bob's PBS is $\alpha |H\rangle_s + \beta |V\rangle _s$.
Photons are sent from Alice to Bob
at a certain interval, 
and the photon in the $i$th time bin of mode $m\in\{S,s,L,l\}$ is 
denoted by $|\cdot \rangle _m^{(i)}$.
Nonlinear interactions for photons 
are intrinsically quite weak in the optical fiber, and the fluctuation of the optical fiber is typically slow.
Therefore, we can model the noise of the optical fiber 
by unknown collective unitary operators $\hat{U}_S$ and $\hat{U}_L$
acting on the polarization qubit in modes $S$ and $L$, respectively.
Since 
the input photon in mode $S$ and $L$ is 
$H$ and $V$ polarized, respectively,
the set of complex parameters $\delta\equiv(a,b,c,d)$ 
defined by 
\begin{eqnarray*}
\hat{U}_S |H\rangle _S = a|H\rangle_S+b|V\rangle _S,\;\;
\hat{U}_L |V\rangle _L = c|H\rangle_L+d|V\rangle _L,\;\;
\end{eqnarray*}
and $|a|^2+|b|^2=|c|^2+|d|^2=1$
is enough to model the collective unitary error of the quantum channel~\cite{[YSOKI05]}.
The parties will extract the DFS 
from photons in 
different time bins,
where we assume that the time difference is sufficiently 
small compared to the fluctuation time 
of 
$\delta$.

The 
BFK protocol runs as follows~\cite{[BFK09]}: 
(1) Alice sends $n$ rotated qubits $\{|+_{\theta_j}\rangle\equiv(|0\rangle+e^{i\theta_j}|1\rangle)/\sqrt{2}\}_{j=1}^n$ to Bob. Here, 
$\theta_j$ is randomly chosen such that $\theta_j\in\{{k\pi}/4|k\in\mathbb{Z},0\le k\le7\}$.
(2) Bob generates 
a brickwork state
according to Alice's instruction $\bf{M}$, 
which tells the parties 
how the brickwork state is generated 
from the rotated qubits. 
(3) Bob performs the measurement on the $j$th qubit
according to the measurement angle 
$\xi_j=\theta_j+\phi'_j+r_j\pi$
sent from Alice.
Here, $\phi' _j$ is the angle by which Alice wants to perform the measurement,
and $r_j\in\{0,1\}$ is a random number.
Bob sends the measurement outcome to Alice.
Alice and Bob repeat these procedures to complete MBQC.

Two essential properties of the BQC protocols  
are correctness and blindness.
A BQC protocol is correct 
if and only if the output 
of the protocol is Alice's desired one
as long as Alice and Bob follow
 the procedure of the protocol faithfully.
A protocol is blind 
if and only if Bob cannot know any information about Alice's inputs, algorithm, and outputs whenever Alice follows the procedure of the protocol.

\medskip
\begin{center}
\bf{III. ENTANGLEMENT-BASED PROTOCOL}
\end{center}

The first DFS-BQC protocol runs as follows: 
(1) Alice generates the Bell pair 
$|\Psi^+\rangle^{(i)}\equiv(|H\rangle^{(2i-1)}|V\rangle^{(2i)}+|V\rangle^{(2i-1)}|H\rangle^{(2i)})/\sqrt{2}$, 
which is in the DFS against the collective dephasing. 
(2) Alice randomly rotates 
the $(2i-1)$th photon of $|\Psi^+\rangle^{(i)}$ by
$\hat{R}_z(\theta_i)\equiv e^{-i\frac{\theta_i}{2}\hat{Z}}$ ($\theta_i\in\{{k\pi}/4|k\in\mathbb{Z},0\le k\le7\}$).
Alice sends the rotated Bell pair 
to Bob by using the quantum channel. 
(3) Bob performs the quantum nondemolition (QND) measurement 
of the photon numbers $N_{m}^{(2i-1,2i)} \equiv n^{(2i-1)}_{m}+n^{(2i)}_{m}$,
where $n^{(k)}_{m}$ indicates the 
photon number 
of the $k$th time bin and the spatial mode $m$~\cite{[IHY85]}.
When the 
outcome of the QND measurement is ($N^{(2i-1,2i)}_{s}, N^{(2i-1,2i)}_{l})=(2,0), (0,2)$, or $(1,1)$,
Bob tells Alice that the $i$th Bell pair 
is successfully sent from Alice to Bob.
According to the measurement outcome, Bob performs an appropriate
operation
so as to extract 
the signal qubit protected by the DFS, 
and they proceed to step (4).
When Bob obtains other measurement outcomes,
this protocol fails, 
and  
they return to step (1).
(4) Alice and Bob repeat steps (1)--(3)
until these steps succeed $n$ times. 
(5) 
The remaining steps are the same as steps (2) and (3) of the 
BFK protocol~\cite{[BFK09]}.

Below,
we will show correctness.
\\
{\it Proof:}
The QND measurement in step (3) eliminates the effect of the photon loss, 
and hence we 
consider only cases in which two photons arrive on Bob's side. 
The state after Bob's PBS is
$\{(a|H\rangle_s^{(2i-1)}+b|V\rangle_l^{(2i-1)})(c|H\rangle_l^{(2i)}+d|V\rangle_s^{(2i)})
+e^{i\theta_i}(c|H\rangle_l^{(2i-1)}+d|V\rangle_s^{(2i-1)})(a|H\rangle_s^{(2i)}+b|V\rangle_l^{(2i)})\}/\sqrt{2}$.
There are three
successful cases 
according to Bob's measurement outcomes
in step (3) as follows:
(i) In the case of $(N^{(2i-1,2i)}_{s},N^{(2i-1,2i)}_{l})=(2,0)$,
the state is projected to 
$(|H\rangle_s^{(2i-1)}|V\rangle_s^{(2i)}+e^{i\theta_i}|V\rangle_s^{(2i-1)}|H\rangle_s^{(2i)})/\sqrt{2}$
with probability ${|ad|}^2T^2$.
(ii) In the case of $(N^{(2i-1,2i)}_{s},N^{(2i-1,2i)}_{l})=(0,2)$,
the state is projected to 
$(|H\rangle_l^{(2i-1)}|V\rangle_l^{(2i)}+e^{-i\theta_i}|V\rangle_l^{(2i-1)}|H\rangle_l^{(2i)})/\sqrt{2}$
with probability ${|bc|}^2T^2$. 
(iii) In the case of $(N^{(2i-1,2i)}_{s},N^{(2i-1,2i)}_{l})=(1,1)$,
the state is projected to $\{ac(|H\rangle_s^{(2i-1)}|H\rangle_l^{(2i)}+e^{i\theta_i}|H\rangle_l^{(2i-1)}|H\rangle_s^{(2i)})+bd(|V\rangle_l^{(2i-1)}|V\rangle_s^{(2i)}+e^{i\theta_i}|V\rangle_s^{(2i-1)}|V\rangle_l^{(2i)})\}/\sqrt{2(|ac|^2+|bd|^2)}$ with probability $({|ac|}^2+{|bd|}^2)T^2$. 
In any cases,
Bob obtains $|\Psi _{\theta _i}\rangle_s^{(i)}\equiv(|H\rangle
_s^{(2i-1)}|V\rangle_s^{(2i)}+e^{i\theta_i}|V\rangle_s^{(2i-1)}|H\rangle_s^{(2i)})/\sqrt{2}$
by applying an appropriate 
operation
according to the measurement outcome. Particularly in case (iii), where two photons are detected on both modes, Bob transforms the state by using $\hat{U}_{\rm{p.f.}}\equiv|H\rangle_s\langle H|_s+|H\rangle_l\langle V|_s+|V\rangle_s\langle H|_l+|V\rangle_l\langle V|_l$ and performs the QND measurement again.
Finally, Bob performs the CNOT gate $\hat{\Lambda}(\hat{X})$ to $|\Psi _{\theta _i}\rangle_s^{(i)}$
to obtain 
the desired qubit
$|+_{\theta_i}\rangle$ as the first (control) qubit. 
This state is the same as the state of the rotated qubit in step (1) of the BFK protocol (see Appendix A).\hspace{\fill}$\blacksquare$

The total success probability becomes
$T^2$,
which means that Bob deterministically obtains the
 desired qubit up to the photon loss.

Next,
we will show blindness.
\\
{\it Proof:}
The information sent from Alice to Bob 
is $\hat{U}_{\rm{PBS}}|\Psi_{\theta_i}\rangle^{(i)}$, $n$, $\bf{M}$, 
and 
$\xi_i$,
where
$\hat{U}_{\rm{PBS}}$ represents 
the operation by the PBS. 
In addition, according to the message $\bf{m}$ 
that
tells 
whether or not the protocol succeeds, 
Alice needs to decide whether or not she sends additional Bell pairs. 
Since the measurement angle $\xi_i$
is shifted by $r_i\pi$ with a random variable $r_i\in\{0,1\}$,
the quantum state sent from Alice to Bob 
is written, from Bob's point of view, 
as
\begin{eqnarray*}
&&\bigotimes_{i=1}^{N'}\sum_{r_i=0}^1 \frac{1}{2}\hat{U}_{\rm{PBS}}|\Psi_{\theta_i}\rangle^{(i)}\langle\Psi_{\theta_i}|^{(i)}\hat{U}_{\rm{PBS}}^{\dag}\\
&&=\bigotimes_{i=1}^{N'}\hat{U}_{\rm{PBS}}\hat{\Lambda}(\hat{X})({\frac{\hat{I}}{2}}^{(2i-1)}\otimes|V\rangle^{(2i)}\langle V|^{(2i)})\hat{\Lambda}(\hat{X})\hat{U}_{\rm{PBS}}^{\dag}.
\end{eqnarray*}
Here, $N'$ is the actual number of Bell pairs sent from Alice to Bob, and depends on only $n$ and $\bf{m}$. 
The above state does not 
depend on 
any information about 
Alice's inputs, algorithm, and outputs,
and hence the entanglement-based DFS-BQC protocol 
has blindness.\hspace{\fill}$\blacksquare$

\medskip
\begin{center}
\bf{IV. SINGLE-PHOTON-BASED PROTOCOL}
\end{center}

We propose a single-photon-based DFS-BQC protocol,
where
the extraction of the DFS against the collective dephasing (DFS extraction, DFSE) is utilized
in order to remove the necessity of the entanglement generation on Alice's side~\cite{[YSOKI05]}.
In this protocol, all Alice has 
to do is the same as what she has to do in the BFK protocol.

The DFSE for two photons proceeds as follows: 
(1) Apply 
$\hat{\Lambda}(\hat{X})$ to 
two photons.
(2) Measure the target qubit in the $\hat{Z}$ basis.
If the outcome implies $|V\rangle$, then
the remaining control qubit comes from the DFS,
and the DFSE is successfully done. Otherwise, the DFSE fails.

The single-photon-based DFS-BQC
runs as follows:
(1) Alice generates $2N$ rotated photons $\{|+_{\theta_i}\rangle^{(i)}\}_{i=2N(h-1)+1}^{2Nh}$, and sends them to Bob by using the quantum channel.
Here, 
$\theta_i$ is  
chosen randomly, similarly to the previous case, and $h$ is the number of the repetition of steps (1)--(4). 
The number of photons $2N$ is chosen such that all $2N$ photons experience 
the collective noise.
In other words, $N$ is determined by 
the time scale of the fluctuation 
of the optical fiber and the repetition rate of the single-photon source.
(2) Bob performs the QND measurement of the photon number 
$n^{(i)}_{s}+n^{(i)}_{l}$.
Bob constructs $\lfloor M/2\rfloor$ pairs of the photons in $k$th and $k'$th time bins with $n_s^{(k)}+n_l^{(k)}=1$ and $n_s^{(k')}+n_l^{(k')}=1$, where $M$ is the total number of time bins satisfying $n_s^{(i)}+n_l^{(i)}=1$.
(3) Bob performs the QND measurement of the photon number $N^{(k,k')}_{m}$.
If $(N^{(k,k')}_{s},N^{(k,k')}_{l})=(2,0)$ or $(0,2)$, they proceed to step (4). 
On the other hand, if $(N^{(k,k')}_{s},N^{(k,k')}_{l})=(1,1)$, 
Bob performs $\hat{U}_{\rm{p.f.}}\otimes\hat{U}_{\rm{p.f.}}$ 
to the output. 
Then, he performs the QND measurement of the photon number $N^{(k,k')}_{m}$ 
again.
If the outcome of the second QND measurement satisfies $(N^{(k,k')}_{s},N^{(k,k')}_{l})=(2,0)$ or $(0,2)$, they proceed to step (4).
If the outcome of the second QND measurement satisfies that $(N^{(k,k')}_{s},N^{(k,k')}_{l})=(1,1)$, they discard the pair and perform step (3) over again.
If all $\lfloor M/2\rfloor$ pairs are consumed, 
they return to step (1).
(4) Bob performs the DFSE 
for the pair.
If the DFSE succeeds for the $k$th and $k'$th photons, 
$|+_{\theta_k-\theta_{k'}}\rangle$ is obtained, 
and Bob tells Alice that it succeeds. 
If the DFSE fails, they return to step (3)
to obtain another pair.
(5) Alice and Bob repeat (1)--(4) until these steps succeed $n$ times. 
(6) The remaining steps are the same as steps (2) and (3) of the 
BFK protocol.

The correctness of this protocol 
is proven
in the same way as the entanglement-based DFS-BQC protocol,
except that the extraction of
the desired qubit
becomes probabilistic. 
Since the success probability of the DFSE is $1/2$,
the probability of obtaining the desired qubit 
from 
$2N$ photons 
is calculated to be $\sum_{M=0}^{2N}\binom{2N}{M}T^M{(1-T)}^{2N-M}(1-1/2^{\lfloor M/2\rfloor})$,
which rapidly approaches to unity for 
sufficiently large $N$, as shown in Fig.~\ref{comparison} (see Appendix B).
\begin{figure}[t]
 \begin{center}
  \includegraphics[width=8.6cm,clip]{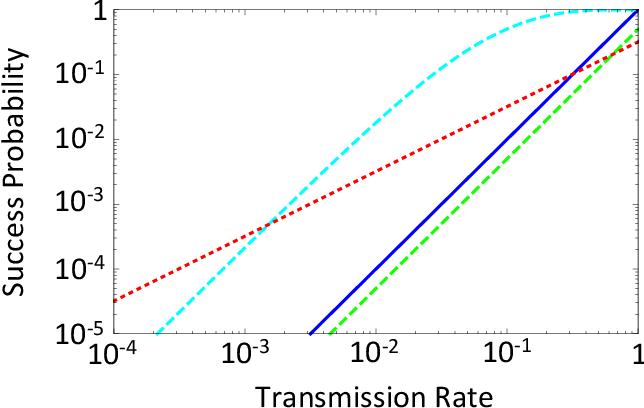}
\end{center}
\captionsetup{justification=raggedright,singlelinecheck=false}
\caption{The success probabilities
for the entanglement-based 
(blue line),
the single-photon-based 
(cyan dashed line for $N=15$ and green dashed line for $N=1$),
and the coherent-light-assisted (red dotted line for $\mu=1/T$ and $|a|=|d|=1/\sqrt{2}$) DFS-BQC protocol (see Appendix D).}
\label{comparison}
\end{figure}

Next, we will show blindness.
\\
{\it Proof:}
The quantum state sent from Alice to Bob
is written, from Bob's point of view, 
as
\begin{eqnarray*}
\bigotimes_{i=1}^{N'}\sum_{{r_i}=0}^{1}\frac{1}{2}\hat{U}_{\rm{PBS}}|+_{\theta_i}\rangle^{(i)}\langle+_{\theta_i}|^{(i)}\hat{U}_{\rm{PBS}}^\dag
=\bigotimes_{i=1}^{N'}\hat{U}_{\rm{PBS}}\frac{\hat{I}}{2}^{(i)}\hat{U}_{\rm{PBS}}^\dag.
\end{eqnarray*}
Here, $N'$ is the actual number of photons sent from Alice to Bob, and depends on only $N$, $n$, and $\bf{m}$. 
The above state does not 
depend on 
any information about 
Alice's inputs, algorithm, and outputs,
and hence the single-photon-based DFS-BQC protocol 
has blindness.\hspace{\fill}$\blacksquare$

\medskip
\begin{center}
\bf{V. COHERENT-LIGHT-ASSISTED PROTOCOL}
\end{center}

The success probability of the single-photon-based DFS-BQC protocol scales 
$O(T^2)$. 
If the quantum channel is very lossy, the success probability of this protocol becomes very low.
In order to improve the efficiency, 
we propose a coherent-light-assisted DFS-BQC protocol~\cite{[IOTYKI11],[KYKI13]}.

The coherent-light-assisted DFS-BQC protocol runs as follows:
(1) Alice generates a rotated photon $|+_{\theta_i}\rangle^{(2i-1)}$ and a coherent-light pulse 
$|\mu\rangle_+^{(2i)}\equiv e^{-\mu/2}\sum_{n'=0}^{\infty}{(\sqrt{\mu})}^{n'}/\sqrt{{n'}!}|n'\rangle_{+}^{(2i)}$,
where the subscript $+$ indicates  
the polarization $|+\rangle\equiv|+_0\rangle$ and the phase of the coherent light is fixed to $0$ for simplicity. $\theta_i$ is chosen randomly, similarly to the previous case.
She sends these two states alternately to Bob by using the quantum channel. 
(2) Bob performs the QND measurements of the photon number $n_s^{(2i-1)}+n_l^{(2i-1)}$ and $n_s^{(2i)}+n_l^{(2i)}$ for the $(2i-1)$th photon and the $2i$th coherent light, 
respectively. 
If any of the events with $n_s^{(2i-1)}+n_l^{(2i-1)}=1$ 
and 
$n_s^{(2i)}+n_l^{(2i)}\ge1$
are obtained, 
they proceed to step (3). 
Otherwise,
they return to step (1). 
(3) Bob performs almost the same procedure as step (3) of the single-photon-based protocol with $M=2$ for the $(2i-1)$th photon and a photon which is extracted from the $2i$th Fock state. Unlike step (3) of the single-photon-based protocol, if the outcome of the second QND measurement satisfies that $(N_s^{(2i-1,2i)},N_l^{(2i-1,2i)})=(1,1)$, he measures the mode of the $2i$th photon nondestructively and flips the polarization of the $2i$th photon. After that, they proceed to step (4).
(4) Bob performs the DFSE for the pair of $(2i-1)$th and $2i$th photons. If the DFSE succeeds, $|+_{\theta_i}\rangle$ is obtained, and he tells Alice that it succeeds. If the DFSE fails, they proceed to step (5). (5) Bob performs the QND measurement of the photon number $n_m^{(2i)}$. According to the outcome(s) in step (3), he discards the photon in mode $l$, $s$, or does nothing. Then, he repeats the DFSEs with the output of the previous DFSE and one of the remaining photons. When the number of the successful DFSEs 
exceeds that of the failure DFSEs, 
$|+_{\theta_i}\rangle$ is obtained, and he tells Alice that it succeeds. 
If all 
remaining photons are consumed, 
they return to step (1). 
(6) Alice and Bob repeat steps (1)--(5) until these steps succeed $n$ times. 
(7) The remaining steps are the same as steps (2) and (3) of the
BFK protocol.

To obtain an intuition of correctness, let us look at the following example case (a rigorous proof of correctness is given in Appendix C).
Alice prepares the state
$|+_{\theta_i}\rangle^{(2i-1)}\otimes|\mu\rangle_+^{(2i)}$,
which is sent to Bob by using the quantum channel.
Suppose Bob obtains $\hat{n}_s^{(2i-1)}+\hat{n}_l^{(2i-1)}=1$ in step(2) and 
$(N_s^{(2i-1,2i)},N_l^{(2i-1,2i)})=(2,0)$ in step (3). Then, the state of two photons becomes $(a|H\rangle_s^{(2i-1)}+e^{i\theta_i}d|V\rangle_s^{(2i-1)})/\sqrt{|a|^2+|d|^2}\otimes(a|H\rangle_s^{(2i)}+d|V\rangle_s^{(2i)})/\sqrt{|a|^2+|d|^2}$. In step (4), if the DFSE fails, the state becomes $(a^2|H\rangle_s^{(2i-1)}+e^{i\theta_i}d^2|V\rangle_s^{(2i-1)})/\sqrt{|a|^4+|d|^4}$. Then, he repeats the DFSEs in step (5). After that, if the DFSE succeeds twice in a row, the state changes as follows:
$(a^2|H\rangle_s^{(2i-1)}+e^{i\theta_i}d^2|V\rangle_s^{(2i-1)})/\sqrt{|a|^4+|d|^4}
\rightarrow 
(a|H\rangle_s^{(2i-1)}+e^{i\theta_i}d|V\rangle_s^{(2i-1)})/\sqrt{|a|^2+|d|^2}
\rightarrow 
(|H\rangle_s^{(2i-1)}+e^{i\theta_i}|V\rangle_s^{(2i-1)})/\sqrt{2}=|+_{\theta_i}\rangle_s^{(2i-1)}$.

The success probability 
is calculated by considering a random walk with an absorbing boundary.  When the mean photon number of the coherent light
as an ancilla $\mu$ is $1/T$, the success probability becomes 
$O(T)$, as shown in Fig.~\ref{comparison}. (The derivation of the success probability is given in Appendix D.) Similar to the single-photon-based DFS-BQC protocol, the success probability of the coherent-light-assisted DFS-BQC protocol can be improved by increasing the number of pairs of the single photon and the coherent-light pulse as long as they experience the collective noise. The above protocol is done using one of the single photons and one of the coherent-light pulses or another one of the single photons among them.

Next, we will show blindness.\\
{\it Proof:}
The information sent from Alice to Bob 
is $\hat{U}_{\rm{PBS}}|+_{\theta_i}\rangle^{(2i-1)}$, $\hat{U}_{\rm{PBS}}|\mu\rangle_+^{(2i)}$, $n$, $\bf{M}$, and $\xi_i$. 
The quantum state sent from Alice to Bob is written, from Bob's point of view, as
\begin{eqnarray*}
&&\bigotimes_{i=1}^{N'}\sum_{r_i=0}^{1}\frac{1}{2}\hat{U}_{\rm{PBS}}(|+_{\theta_i}\rangle^{(2i-1)}\langle+_{\theta_i}|^{(2i-1)}\otimes|\mu\rangle_+^{(2i)}\langle\mu|_+^{(2i)}){\hat{U}_{\rm{PBS}}^{\dag}}\\
&&=\bigotimes_{i=1}^{N'}\hat{U}_{\rm{PBS}}(\frac{\hat{I}}{2}^{(2i-1)}\otimes|\mu\rangle_+^{(2i)}\langle\mu|_+^{(2i)})\hat{U}_{\rm{PBS}}^{\dag}.
\end{eqnarray*}
Here, $N'$ is the actual number of photons sent from Alice to Bob, and depends on only $n$ and $\bf{m}$. The above state does not
depend on any information about Alice's inputs, algorithm, and outputs, and hence the coherent-light-assisted DFS-BQC protocol has blindness.\hspace{\fill}$\blacksquare$

\medskip
\begin{center}
\bf{VI. CONCLUSION}
\end{center}

We have proposed 
three kinds of 
DFS-BQC protocols, 
which tolerate 
the collective noise in the 
quantum 
channel. 
While we 
have considered the BFK protocol only, 
it is straightforward to apply 
the proposed DFS-based schemes 
for other BQC protocols, 
such as the topological BQC protocol~\cite{[MF12]} 
and unconditionally verifiable BQC protocol~\cite{[FK12]},
without degrading their unconditional security. 
Moreover, while we have assumed the collective unitary noise and loss as the imperfection of the 
quantum channel, it is straightforward to extend the proposed protocols to more general collective noise with collective Kraus operators.

\medskip
\begin{center}
\bf{ACKNOWLEDGMENTS}
\end{center}

We thank Y. Nagamatsu
 for helpful discussions. This work was supported by Program for Leading Graduate Schools: ``Interactive Materials Science Cadet Program," and JSPS KAKENHI Grants No. 25247068, No. 15H03704, and No. 16H02214.

\medskip
\begin{center}
\bf{APPENDIX A: BOB'S OPERATIONS IN 
THE ENTANGLEMENT-BASED DFS-BQC PROTOCOL}
\end{center}

In order to complete the proof of correctness
of the entanglement-based DFS-BQC protocol, here
we explain Bob's operations 
employed after the QND measurements in step (3) of Sec. III.

All operations are designed to obtain 
$|\Psi _{\theta _i} \rangle^{(i)}_s$ 
from the state after the QND measurements.
In case (i) with $(N_s^{(2i-1,2i)},N_l^{(2i-1,2i)}) = (2,0)$,
Bob obtains $|\Psi _{\theta _i} \rangle^{(i)}_s$
and hence no operation is required.
In case (ii) with $(N_s^{(2i-1,2i)},N_l^{(2i-1,2i)}) = (0,2)$,
the state is projected to
\begin{eqnarray*}
\frac{|H\rangle_l^{(2i-1)}|V\rangle_l^{(2i)}+e^{-i\theta_i}|V\rangle_l^{(2i-1)}|H\rangle_l^{(2i)}}{\sqrt{2}}.
\end{eqnarray*}
By applying the Pauli-$X$ operation and the swapping operation of modes $l$ and $s$ 
for two photons,
he obtains
\begin{eqnarray*}
\frac{|H\rangle_s^{(2i-1)}|V\rangle_s^{(2i)}+e^{i\theta_i}|V\rangle_s^{(2i-1)}|H\rangle_s^{(2i)}}{\sqrt{2}}=|\Psi_{\theta_i}\rangle^{(i)}_s.
\end{eqnarray*}
In case (iii) with $(N_s^{(2i-1,2i)},N_l^{(2i-1,2i)}) = (1,1)$,
the state is projected to
\begin{eqnarray*}
&&\{ac(|H\rangle_s^{(2i-1)}|H\rangle_l^{(2i)}+e^{i\theta_i}|H\rangle_l^{(2i-1)}|H\rangle_s^{(2i)})+\\
&&bd(|V\rangle_l^{(2i-1)}|V\rangle_s^{(2i)}+e^{i\theta_i}|V\rangle_s^{(2i-1)}|V\rangle_l^{(2i)})\}/{\sqrt{2(|ac|^2+|bd|^2)}}.
\end{eqnarray*}
By applying $\hat{U}_{\rm{p.f.}}\equiv|H\rangle_s\langle H|_s+|H\rangle_l\langle V|_s+|V\rangle_s\langle H|_l+|V\rangle_l\langle V|_l$, 
he obtains 
\begin{eqnarray*}
&&\{ac(|H\rangle_s^{(2i-1)}|V\rangle_s^{(2i)}+e^{i\theta_i}|V\rangle_s^{(2i-1)}|H\rangle_s^{(2i)})+\\
&&bd(|V\rangle_l^{(2i-1)}|H\rangle_l^{(2i)}+e^{i\theta_i}|H\rangle_l^{(2i-1)}|V\rangle_l^{(2i)})\}/{\sqrt{2(|ac|^2+|bd|^2)}}.
\end{eqnarray*}
Then, 
he further performs the QND measurement of the photon number $N_m^{(2i-1,2i)}$, 
and always obtains $(N_s^{(2i-1,2i)},N_l^{(2i-1,2i)})=(2,0)$ or $(0,2)$. 
The former case is the same as case (i), 
and the latter case is the same as case (ii). 
Accordingly, he obtains $|\Psi_{\theta_i}\rangle_s^{(i)}$ in any cases.

\medskip
\begin{center}
\bf{
APPENDIX B: CORRECTNESS OF THE SINGLE-PHOTON-BASED DFS-BQC PROTOCOL}
\end{center}

Here we provide a 
detail of the proof of correctness of the single-photon-based DFS-BQC protocol,
where its success probability is also calculated.
\\
{\it Proof:}
The state of the $k$th and $k'$th photons before 
QND measurements
is given by
\begin{eqnarray*}
&&\frac{(a|H\rangle_s^{(k)}+b|V\rangle_l^{(k)})+e^{i\theta_k}(c|H\rangle_l^{(k)}+d|V\rangle_s^{(k)})}{\sqrt{2}}\\
&\otimes&\frac{(a|H\rangle_s^{(k')}+b|V\rangle_l^{(k')})+e^{i\theta_{k'}}(c|H\rangle_l^{(k')}+d|V\rangle_s^{(k')})}{\sqrt{2}}.
\end{eqnarray*}
There are four successful cases 
depending on the outcomes of the 
QND measurements, as follows:\\

\begin{enumerate}[(i)]
\item The first QND measurement in step (3) of Sec. IV results in $(N_s^{(k,k')},N_l^{(k,k')})=(2,0)$.
The state is projected to
\begin{eqnarray*}
&&(a^2|H\rangle_s^{(k)}|H\rangle_s^{(k')}+e^{i\theta_{k'}}ad|H\rangle_s^{(k)}|V\rangle_s^{(k')}+\\
&&e^{i\theta_k}ad|V\rangle_s^{(k)}|H\rangle_s^{(k')}+e^{i(\theta_k+\theta_{k'})}d^2|V\rangle_s^{(k)}|V\rangle_s^{(k')})/(|a|^2+|d|^2)
\end{eqnarray*}
with probability $({|a|}^2+{|d|}^2)^2/4$.
If the DFSE succeeds,
then Alice's desired qubit 
$(|H\rangle+e^{i(\theta_k-\theta_{k'})}|V\rangle)/\sqrt{2}$
 is prepared.
The success probability of the DFSE is calculated to be
$2{|ad|}^2/{({|a|}^2+{|d|}^2)}^2$. 

\item The first QND measurement in step (3) results in $(N_s^{(k,k')},N_l^{(k,k')})=(0,2)$.
The state is projected to
\begin{eqnarray*}
&&(b^2|V\rangle_l^{(k)}|V\rangle_l^{(k')}+e^{i\theta_{k'}}bc|V\rangle_l^{(k)}|H\rangle_l^{(k')}+\\
&&e^{i\theta_k}bc|H\rangle_l^{(k)}|V\rangle_l^{(k')}+e^{i(\theta_k+\theta_{k'})}c^2|H\rangle_l^{(k)}|H\rangle_l^{(k')})/(|b|^2+|c|^2)
\end{eqnarray*}
with probability $({|b|}^2+{|c|}^2)^2/4$. 
If the DFSE succeeds, 
$(|H\rangle+e^{i(\theta_{k'}-\theta_k)}|V\rangle)/\sqrt{2}$ is prepared. 
The Pauli-$X$ operation is applied 
in order to flip the sign of the phase, and $(|H\rangle+e^{i(\theta_k-\theta_{k'})}|V\rangle)/\sqrt{2}$ is obtained. 
The success probability of the DFSE is calculated to be
$2{|bc|}^2/{({|b|}^2+{|c|}^2)}^2$. 

\item 
The first QND measurement in step (3) results in $(N_s^{(k,k')},N_l^{(k,k')})=(1,1)$.
The state is projected to 
\begin{eqnarray*}
&&ab|H\rangle_s^{(k)}|V\rangle_l^{(k')}+e^{i\theta_{k'}}ac|H\rangle_s^{(k)}|H\rangle_l^{(k')}+\\
&&ab|V\rangle_l^{(k)}|H\rangle_s^{(k')}+e^{i\theta_{k'}}bd|V\rangle_l^{(k)}|V\rangle_s^{(k')}+\\
&&e^{i\theta_k}ac|H\rangle_l^{(k)}|H\rangle_s^{(k')}+e^{i(\theta_k+\theta_{k'})}cd|H\rangle_l^{(k)}|V\rangle_s^{(k')}+\\
&&e^{i\theta_k}bd|V\rangle_s^{(k)}|V\rangle_l^{(k')}+e^{i(\theta_k+\theta_{k'})}cd|V\rangle_s^{(k)}|H\rangle_l^{(k')}
\end{eqnarray*}
up to normalization. By applying $\hat{U}_{\rm{p.f.}}\otimes\hat{U}_{\rm{p.f.}}$,
Bob obtains
\begin{eqnarray*}
&&ab|H\rangle_s^{(k)}|V\rangle_l^{(k')}+e^{i\theta_{k'}}ac|H\rangle_s^{(k)}|V\rangle_s^{(k')}+\\
&&ab|V\rangle_l^{(k)}|H\rangle_s^{(k')}+e^{i\theta_{k'}}bd|V\rangle_l^{(k)}|H\rangle_l^{(k')}+\\
&&e^{i\theta_k}ac|V\rangle_s^{(k)}|H\rangle_s^{(k')}+e^{i(\theta_k+\theta_{k'})}cd|V\rangle_s^{(k)}|H\rangle_l^{(k')}+\\
&&e^{i\theta_k}bd|H\rangle_l^{(k)}|V\rangle_l^{(k')}+e^{i(\theta_k+\theta_{k'})}cd|H\rangle_l^{(k)}|V\rangle_s^{(k')}
\end{eqnarray*}
up to normalization. Bob performs the QND measurement of the photon number $N^{(k,k')}_{m}$ again.
There are two 
successful cases in case (iii), as follows:
\begin{itemize}
\item[(iii-i)]
$(N^{(k,k')}_s,N^{(k,k')}_l)=(2,0).$
The state is projected to
\begin{eqnarray*}
\frac{|H\rangle_s^{(k)}|V\rangle_s^{(k')}+e^{i(\theta_k-\theta_{k'})}|V\rangle_s^{(k)}|H\rangle_s^{(k')}}{\sqrt{2}}.
\end{eqnarray*}
The DFSE for this state
always succeeds, 
and $(|H\rangle+e^{i(\theta_k-\theta_{k'})}|V\rangle)/\sqrt{2}$ is prepared. 
The overall success probability of the present case is calculated to be $|ac|^2/2$.

\item[(iii-ii)]
$(N^{(k,k')}_{s},N^{(k,k')}_{l})=(0,2)$.
The state is projected to
\begin{eqnarray*}
\frac{|H\rangle_l^{(k)}|V\rangle_l^{(k')}+e^{i(\theta_{k'}-\theta_k)}|V\rangle_l^{(k)}|H\rangle_l^{(k')}}{\sqrt{2}}.
\end{eqnarray*}
The DFSE for this state always succeeds, 
and $(|H\rangle+e^{i(\theta_{k'}-\theta_k)}|V\rangle)/\sqrt{2}$ is prepared. 
The Pauli-$X$ operation is applied 
in order to flip the sign of the phase, and $(|H\rangle+e^{i(\theta_k-\theta_{k'})}|V\rangle)/\sqrt{2}$ is obtained. 
The overall success probability of the present case is calculated to be $|bd|^2/2$.
\end{itemize}
\end{enumerate}
Accordingly, if the DFSE succeeds, 
$|+_{\theta_k-\theta_{k'}}\rangle$ is prepared on Bob's side.\hspace{\fill}$\blacksquare$

We derive the probability 
of the successful preparation of Alice's desired 
qubit with 
$2N$ single photons,
which experience the same unitary noise and the photon loss. 
First,
we calculate the success probability of the DFSE for a pair of two photons. 
This is done by summing all 
success probabilities shown in the above proof of correctness:
\begin{eqnarray*}
&&\frac{(|a|^2+|d|^2)^2}{4}\frac{2|ad|^2}{(|a|^2+|d|^2)^2}+\frac{(|b|^2+|c|^2)^2}{4}\frac{2|bc|^2}{(|b|^2+|c|^2)^2}+\\
&&\frac{|ac|^2}{2}+\frac{|bd|^2}{2}=\frac{1}{2}.
\end{eqnarray*}
This indicates that 
the net failure probability of the DFSE for each pair of two photons is $1/2$. 
Since Alice sends $2N$ photons by using the quantum channel with the transmission rate $T$,
the probability 
that $M$ photons reach Bob's side is 
calculated to be $\binom{2N}{M}T^M{(1-T)}^{2N-M}$. 
Since the DFSE is done for each pair of two photons independently, 
the success probability of this protocol 
is given by
\begin{eqnarray*}
\sum_{M=0}^{2N}\binom{2N}{M}T^M{(1-T)}^{2N-M}{\biggl(}1-\frac{1}{2^{\lfloor\frac{M}{2}\rfloor}}{\biggl)}.
\end{eqnarray*}

\medskip
\begin{center}
\bf{APPENDIX C: CORRECTNESS OF THE COHERENT-LIGHT-ASSISTED DFS-BQC PROTOCOL}
\end{center}

Here we provide the proof of correctness of the coherent-light-assisted DFS-BQC protocol.
\\
{\it Proof:}
From correctness of the single-photon-based DFS-BQC protocol, it is proven that if the DFSE succeeds in step (4) of Sec. V, a desired qubit is prepared. Thus, without loss of generality, we consider only the case that Alice and Bob proceed to step (5). In order to prove correctness, we have to consider three cases depending on the outcome(s) of the QND measurement(s) in step (3).\\

\begin{enumerate}[(i)]
\item The first QND measurement in step (3) results in $(N_s^{(2i-1,2i)},N_l^{(2i-1,2i)})=(2,0)$, and the DFSE fails in step (4). First, Bob discards the photons that exist in mode $l$. Bob repeats the same procedure as step (4), that is, the DFSE for the output of the previous DFSE and one of the remaining photons extracted from the coherent light. Suppose the DFSEs succeed and fail $N_{\rm{right}}$($\ge 0$) and $N_{\rm{left}}$($\ge 1$) times, respectively. In such a case, the state is transformed into
\begin{eqnarray*}
a^{1+N_{\rm{left}}}d^{N_{\rm{right}}}|H\rangle_s^{(2i-1)}+e^{i\theta_i}a^{N_{\rm{right}}}d^{1+N_{\rm{left}}}|V\rangle_s^{(2i-1)}
\end{eqnarray*}
up to normalization. If $N_{\rm{right}}=1+N_{\rm{left}}$ is satisfied, the above state becomes $(|H\rangle_s^{(2i-1)}+e^{i\theta_i}|V\rangle_s^{(2i-1)})/\sqrt{2}=|+_{\theta_i}\rangle_s^{(2i-1)}$. In other words, when the number of the successful DFSEs exceeds that of the failure DFSEs, $|+_{\theta_i}\rangle$ is obtained.

\item The first QND measurement in step (3) results in $(N_s^{(2i-1,2i)},N_l^{(2i-1,2i)})=(0,2)$, and the DFSE fails in step (4). First, Bob discards the photons existing in mode $s$. Bob repeats the DFSE, similarly to the above case (i). Suppose the DFSEs succeed and fail $N_{\rm{right}}$($\ge 0$) and $N_{\rm{left}}$($\ge 1$) times, respectively. In such a case, the state is transformed into
\begin{eqnarray*}
b^{1+N_{\rm{left}}}c^{N_{\rm{right}}}|H\rangle_l^{(2i-1)}+e^{-i\theta_i}b^{N_{\rm{right}}}c^{1+N_{\rm{left}}}|V\rangle_l^{(2i-1)}
\end{eqnarray*}
up to normalization. If $N_{\rm{right}}=1+N_{\rm{left}}$ is satisfied, the above state becomes $(|H\rangle_l^{(2i-1)}+e^{-i\theta_i}|V\rangle_l^{(2i-1)})/\sqrt{2}=|+_{-\theta_i}\rangle_l^{(2i-1)}$. By performing the Pauli-$X$ operation for this state, $|+_{\theta_i}\rangle$ is obtained. In other words, when the number of the successful DFSEs exceeds that of the failure DFSEs, $|+_{\theta_i}\rangle$ is obtained.

\item The first and second QND measurements in step (3) result in $(N_s^{(2i-1,2i)},N_l^{(2i-1,2i)})=(1,1)$. The output of the second DFSE is given by 
\begin{eqnarray*}
&&\{ab(|H\rangle_s^{(2i-1)}|V\rangle_l^{(2i)}+|V\rangle_l^{(2i-1)}|H\rangle_s^{(2i)})+\\
&&e^{i\theta_i}cd(|V\rangle_s^{(2i-1)}|H\rangle_l^{(2i)}+|H\rangle_l^{(2i-1)}|V\rangle_s^{(2i)})\}/{\sqrt{2(|ab|^2+|cd|^2)}}.
\end{eqnarray*}
After Bob measures the spatial mode of the $2i$th photon nondestructively and performs the Pauli-$X$ operation for the $2i$th photon, the above state becomes 
\begin{eqnarray*}
\begin{cases}
\displaystyle\frac{ab|V\rangle_l^{(2i-1)}|V\rangle_s^{(2i)}+e^{i\theta_i}cd|H\rangle_l^{(2i-1)}|H\rangle_s^{(2i)}}{\sqrt{|ab|^2+|cd|^2}},\ \rm{or} & \\
\cfrac{ab|H\rangle_s^{(2i-1)}|H\rangle_l^{(2i)}+e^{i\theta_i}cd|V\rangle_s^{(2i-1)}|V\rangle_l^{(2i)}}{\sqrt{|ab|^2+|cd|^2}} &
\end{cases}
\end{eqnarray*}
depending on the mode of the $2i$th photon. In these cases, the DFSE always fails. The outputs of the DFSE for each of these two states are $(ab|V\rangle_l^{(2i-1)}+e^{i\theta_i}cd|H\rangle_l^{(2i-1)})/\sqrt{|ab|^2+|cd|^2}$ and $(ab|H\rangle_s^{(2i-1)}+e^{i\theta_i}cd|V\rangle_s^{(2i-1)})/\sqrt{|ab|^2+|cd|^2}$, respectively. By performing the Pauli-$X$ operation and swapping the mode, the former state is transformed into the latter state. Thus, without loss of generality, the output of step (4) in this case is regarded as the latter state. 
\begin{itemize}
\item[(iii-i)]
Bob repeats the DFSE for the output of the previous DFSE and one of the remaining photons extracted from the coherent light on mode $s$. Suppose such DFSEs succeed and fail $N_{\rm{right}}$($\ge 0$) and $N_{\rm{left}}$($\ge 1$) times, respectively. In such a case, the state is transformed into
\begin{eqnarray*}
a^{N_{\rm{left}}}bd^{N_{\rm{right}}}|H\rangle_s^{(2i-1)}+e^{i\theta_i}a^{N_{\rm{right}}}cd^{N_{\rm{left}}}|V\rangle_s^{(2i-1)}
\end{eqnarray*}
up to normalization. When $N_{\rm{right}}=N_{\rm{left}}$ is satisfied, the above state becomes $(b|H\rangle_s^{(2i-1)}+e^{i\theta_i}c|V\rangle_s^{(2i-1)})/\sqrt{|b|^2+|c|^2}$, and Bob discards all of the remaining photons extracted from the coherent light on mode $s$.  Bob proceeds to step (iii-ii).
\item[(iii-ii)]
Bob repeats the DFSE for the output of the previous DFSE and one of the remaining photons extracted from the coherent light on mode $l$. Suppose such DFSEs succeed and fail $N'_{\rm{right}}$($\ge 0$) and $N'_{\rm{left}}$($\ge 0$) times, respectively. In such a case, the state is transformed into 
\begin{eqnarray*}
b^{1+N'_{\rm{left}}}c^{N'_{\rm{right}}}|H\rangle_s^{(2i-1)}+e^{i\theta_i}b^{N'_{\rm{right}}}c^{1+N'_{\rm{left}}}|V\rangle_s^{(2i-1)}
\end{eqnarray*}
up to normalization. If $N'_{\rm{right}}=1+N'_{\rm{left}}$ is satisfied, the above state becomes $(|H\rangle_s^{(2i-1)}+e^{i\theta_i}|V\rangle_s^{(2i-1)})/\sqrt{2}=|+_{\theta_i}\rangle_s^{(2i-1)}$.
\end{itemize}
\end{enumerate}
Accordingly, if the number of the successful DFSEs exceeds that of the failure DFSEs, $|+_{\theta_i}\rangle$ is prepared on Bob's side.\hspace{\fill}$\blacksquare$

\begin{figure}[t]
\begin{center}
\includegraphics[width=8.6cm, clip]{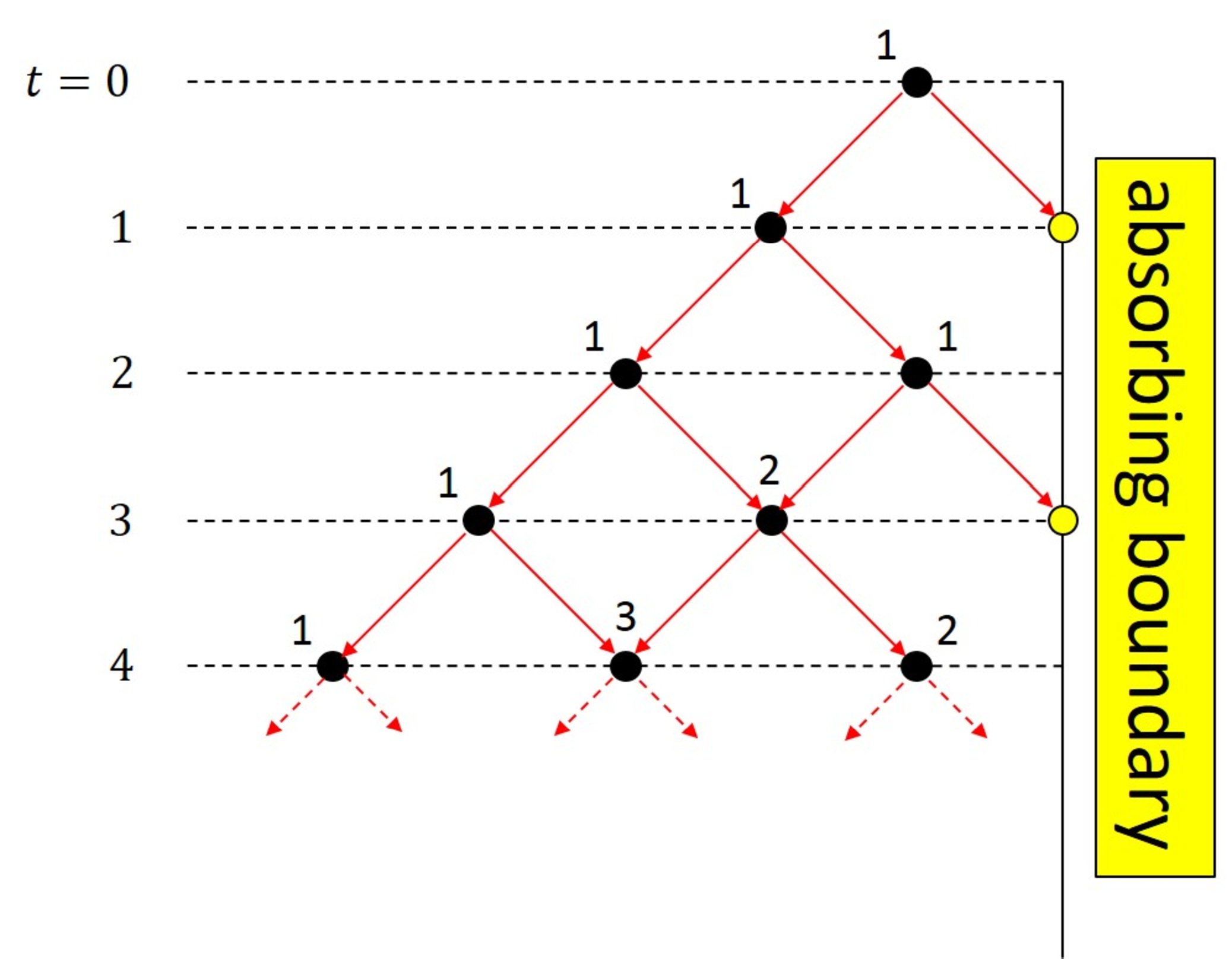}
\end{center}
\captionsetup{justification=raggedright,singlelinecheck=false}
\caption{A classical biased random walk on a line with an absorbing boundary. 
$t$ indicates the number of trials of the DFSEs,
which corresponds to the number of 
steps of the random walk. The numbers
in each site indicate the numbers of paths with which
a walker arrives at that site.}
\label{walk}
\end{figure}

\medskip
\begin{center}
\bf{APPENDIX D: THE $\delta$-DEPENDENCE OF THE SUCCESS PROBABILITY OF THE COHERENT-LIGHT-ASSISTED DFS-BQC PROTOCOL}
\end{center}
Here we derive the success probability of the coherent-light-assisted DFS-BQC protocol 
by using a classical 
biased random walk on a line 
with an absorbing boundary at the right of the starting point.
We regard the successful and failure DFSE
as ``moving right" and ``moving left," respectively, in the classical 
random walk on a line, as shown in Fig.~\ref{walk}.
If the number of 
successful DFSEs exceeds 
that of failure DFSEs, 
a walker 
arrives at the absorbing boundary,
and the protocol finishes successfully.
Let us consider the probability that the walker arrives at the absorbing boundary 
up to a certain time step.
When the walker arrives at the absorbing boundary at the time step $t$, the numbers of moving right $N_{\rm right}(t)$ and moving left  $N_{\rm left}(t)$
have to 
satisfy $N_{\rm right}(t) = N_{\rm left}(t)+1$.
Thus, the walker cannot arrive at the absorbing boundary at an even number step. 
If the walker arrives at the absorbing boundary at time step $(2t'+1)$,
we have $N_{\rm right}(2t'+1) = t'+1$ and $N_{\rm left}(2t'+1)=t'$.
When the walker moves right $N_{\rm{right}}$ times and left $N_{\rm{left}}$ times, in the next step the walker moves right or left with the probability $q_{\Delta N}^{(\cdot)}$ or $(1-q_{\Delta N}^{(\cdot)})$, respectively. These probabilities depend on the cases $(\cdot)\in$\{(i),(ii),(iii-i),(iii-ii)\} in Appendix C and $\Delta N\equiv N_{\rm{left}}-N_{\rm{light}}$. 
Now, we assume that $q_{\Delta N}^{(\cdot)}(1-q_{\Delta N-1}^{(\cdot)})=(1-q_{\Delta N}^{(\cdot)})q_{\Delta N+1}^{(\cdot)}=Q^{(\cdot)}$, which does not depend on $\Delta N$, is satisfied.
In this case, the probability with which 
the random walk is finished
at the time step $(2t'+1)$ 
is given by $C_{t'}q_0^{(\cdot)}{Q^{(\cdot)}}^{t'}$,
where $C_{t'}$ indicates the number of paths 
with which the random walk is finished at the time step $(2t'+1)$. Thus, 
we obtain the total probability 
that the walker arrives at the absorbing boundary
up to time step $t(\ge1)$ as follows:
\begin{eqnarray*}
\sum_{t'=0}^{\lfloor\frac{t-1}{2}\rfloor}C_{t'}q_0^{(\cdot)}{Q^{(\cdot)}}^{t'}.
\end{eqnarray*}

\begin{figure}[t]
 \begin{center}
  \includegraphics[width=8.6cm,clip]{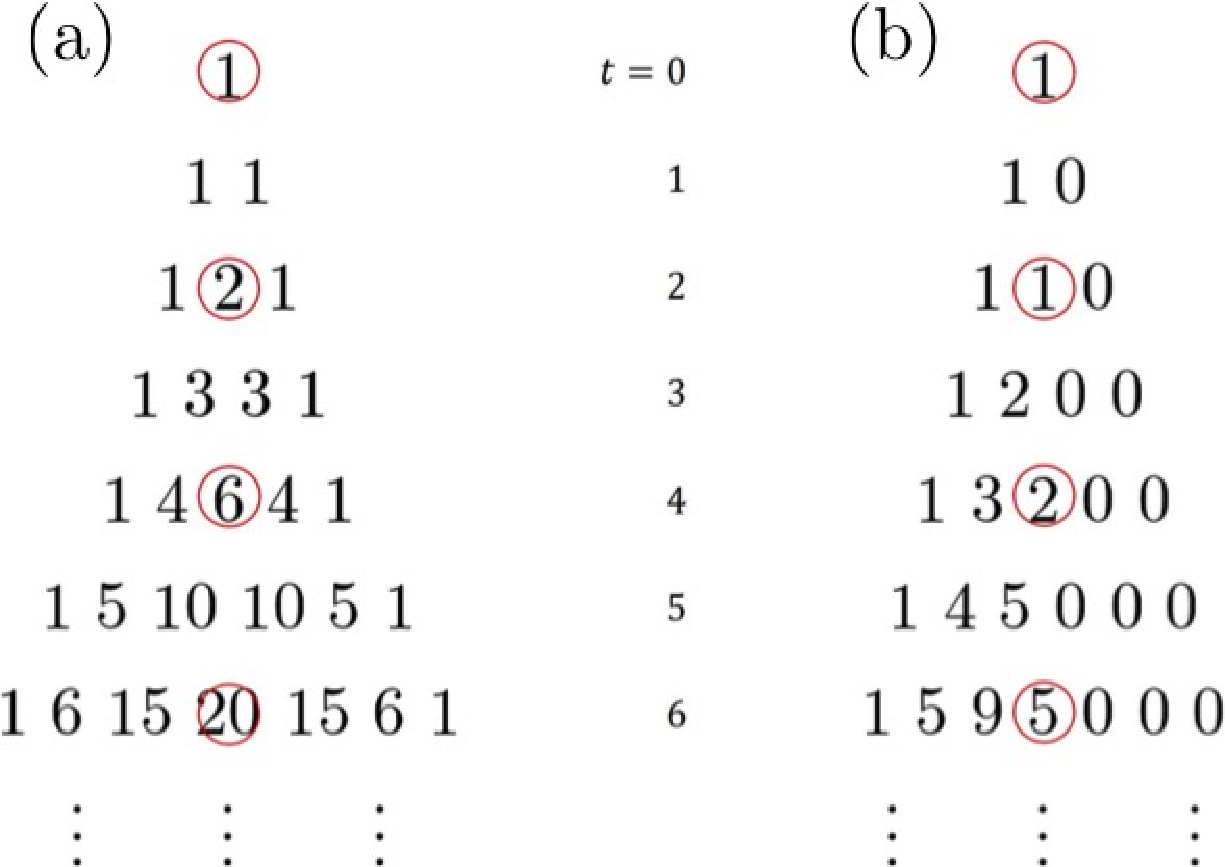}
\end{center}
\captionsetup{justification=raggedright,singlelinecheck=false}
\caption{
The (a) original and (b) modified Pascal's triangles.
Each level $t$ in the triangles corresponds to the number of steps $t$ in Fig.~\ref{walk}. 
The $t$th level numbers circled in red in the modified Pascal's triangle
correspond to the numbers of paths 
of arriving at the absorbing boundary at time step ($t+1$).
}
\label{pascal}
\end{figure}

In order to calculate $C_{t'}$, 
we utilize the original and modified Pascal's triangles, as shown in Fig.~\ref{pascal}. 
The numbers of paths assigned at each node 
correspond to the numbers in the modified Pascal's triangle. 
Because the walker finishes the random walk whenever he arrives at the absorbing boundary, the numbers of paths for the absorbing boundary are written as $0$ as an exception. 
Since the step just before the arrival at the absorbing boundary
is moving right, these numbers are the same as the central numbers of the one step before. That is, $C_{t'}$ corresponds to the $t'$th central number circled in red in the modified Pascal's triangle shown in Fig.~\ref{pascal}(b).
Let $P_{t'}\equiv\binom{2t'}{t'}$ be the $t'$th central number 
of 
the original Pascal's triangle, as shown in Fig.~\ref{pascal}(a). 
By using $P_{t'}$, $C_{t'}$ is given by 
\begin{eqnarray*}
C_{t'}=\frac{P_{t'}}{t'+1}=\frac{\binom{2t'}{t'}}{t'+1}.
\end{eqnarray*}
We prove this equation by using mathematical induction.

{\it Proof.}
We define the $\mathcal{K}$th number in the $t$th level of the original and modified Pascal's triangles as $X_{t,\mathcal{K}}^o$ and $X_{t,\mathcal{K}}^m$, respectively. The first number in the $t$th level indicates the left-most number in the $t$th level. For later convenience, we define $X_{t,0}^o=X_{t,0}^m=0$.

First, we show that $X_{t,\mathcal{K}}^m=X_{t,\mathcal{K}}^o-X_{t,\mathcal{K}-1}^o$ ($1\le\mathcal{K}\le \lceil(t+1)/2\rceil$). From Fig.~\ref{pascal}, this relationship is satisfied for $t=0$, obviously. We assume that this relationship is satisfied for $t=\tau$. Since the properties of the original and modified Pascal's triangle are satisfied for all of $\mathcal{K}$, the equations
\begin{eqnarray*}
X_{\tau+1,\mathcal{K}}^{o(m)}=X_{\tau,\mathcal{K}-1}^{o(m)}+X_{\tau,\mathcal{K}}^{o(m)}
\end{eqnarray*}
are satisfied. This leads to
\begin{eqnarray*}
X_{\tau+1,\mathcal{K}}^m&=&X_{\tau,\mathcal{K}-1}^m+X_{\tau,\mathcal{K}}^m\\
&=&X_{\tau,\mathcal{K}}^o-X_{\tau,\mathcal{K}-2}^0\\
&=&X_{\tau+1,\mathcal{K}}^o-X_{\tau+1,\mathcal{K}-1}^o.
\end{eqnarray*}
By the principle of mathematical induction, we conclude that $X_{t,\mathcal{K}}^m=X_{t,\mathcal{K}}^o-X_{t,\mathcal{K}-1}^o$ ($1\le\mathcal{K}\le \lceil(t+1)/2\rceil$).

Next, we prove that $C_{t'}=P_{t'}/(t'+1)=\binom{2t'}{t'}/(t'+1)$. From Fig.~\ref{pascal}, this relationship is satisfied for $t'=0$, obviously. We assume that $X_{2\tau',\tau'+1}^o=(\tau'+1)X_{2\tau',\tau'+1}^m$ is satisfied for $t'=\tau'$. By using $X_{t,\mathcal{K}}^o=\binom{t}{\mathcal{K}-1}$ as a property of the original Pascal's triangle,
\begin{eqnarray*}
&&X_{2(\tau'+1),\tau'+2}^o\\
&=&2X_{2\tau'+1,\tau'+1}^o\\
&=&2(X_{2\tau',\tau'}^o+X_{2\tau',\tau'+1}^o)\\
&=&2\frac{2\tau'+1}{\tau'+1}X_{2\tau',\tau'+1}^o\\
&=&(\tau'+2)\Biggl\{1-\frac{\tau'(\tau'-1)}{(\tau'+1)(\tau'+2)}\Biggl\}X_{2\tau',\tau'+1}^o\\
&=&(\tau'+2)(X_{2\tau',\tau'+1}^o-X_{2\tau',\tau'-1}^o)\\
&=&(\tau'+2)\Big(\frac{1}{\tau'+1}X_{2\tau',\tau'+1}^o+X_{2\tau',\tau'}^o-X_{2\tau',\tau'-1}^o\Big)\\
&=&(\tau'+2)(X_{2\tau',\tau'+1}^m+X_{2\tau',\tau'}^m)\\
&=&(\tau'+2)X_{2\tau'+1,\tau'+1}^m=(\tau'+2)X_{2(\tau'+1),\tau'+2}^m
\end{eqnarray*}
is satisfied. By the principle of mathematical induction, we conclude that $C_{t'}=P_{t'}/(t'+1)=\binom{2t'}{t'}/(t'+1)$.\hspace{\fill}$\blacksquare$

So far, we have assumed that $q_{\Delta N}^{(\cdot)}(1-q_{\Delta N-1}^{(\cdot)})=(1-q_{\Delta N}^{(\cdot)})q_{\Delta N+1}^{(\cdot)}=Q^{(\cdot)}$ is satisfied. In the following, we prove that this assumption is satisfied in our protocol.

{\it Proof.}
When the walker moves right and left $N_{\rm{right}}$ and $N_{\rm{left}}$ times, the probabilities that the walker moves left the next time in cases (i)--(iii) in Appendix C are given by
\begin{eqnarray*}
&&1-q_{\Delta N}^{(\cdot)}\\
&&=\begin{cases}
\cfrac{|a|^{2(2+\Delta N)}+|d|^{2(2+\Delta N)}}{(|a|^{2(1+\Delta N)}+|d|^{2(1+\Delta N)})(|a|^2+|d|^2)}\ \cdot\cdot\cdot\rm{(i)} & \\
\cfrac{|b|^{2(2+\Delta N)}+|c|^{2(2+\Delta N)}}{(|b|^{2(1+\Delta N)}+|c|^{2(1+\Delta N)})(|b|^2+|c|^2)} \cdot\cdot\cdot\rm{(ii), (iii-ii)} & \\
\cfrac{|a^{1+\Delta N}b|^2+|cd^{1+\Delta N}|^2}{(|a^{\Delta N}b|^2+|cd^{\Delta N}|^2)(|a|^2+|d|^2)}\cdot\cdot\cdot\rm{(iii-i)}. &
\end{cases}
\end{eqnarray*}
Then the values of $Q^{(\cdot)}$ are calculated as
\begin{eqnarray*}
Q^{(\cdot)}=
\begin{cases}
\cfrac{|ad|^2}{(|a|^2+|d|^2)^2}\equiv Q_1\ \cdot\cdot\cdot\rm{(i),\ (iii-i)} & \\
\cfrac{|bc|^2}{(|b|^2+|c|^2)^2}\equiv Q_2\ \cdot\cdot\cdot\rm{(ii),\ (iii-ii)}. &
\end{cases}
\end{eqnarray*}
This means that $q_{\Delta N}^{(\cdot)}(1-q_{\Delta N-1}^{(\cdot)})$ and $(1-q_{\Delta N}^{(\cdot)})q_{\Delta N+1}^{(\cdot)}$ do not depend on $\Delta N$, and the probability that the walker moves right after moving left is exactly the same as that for the walker moving left after moving right.\hspace{\fill}$\blacksquare$

\begin{figure}[t]
 \begin{center}
  \includegraphics[width=8.6cm,clip]{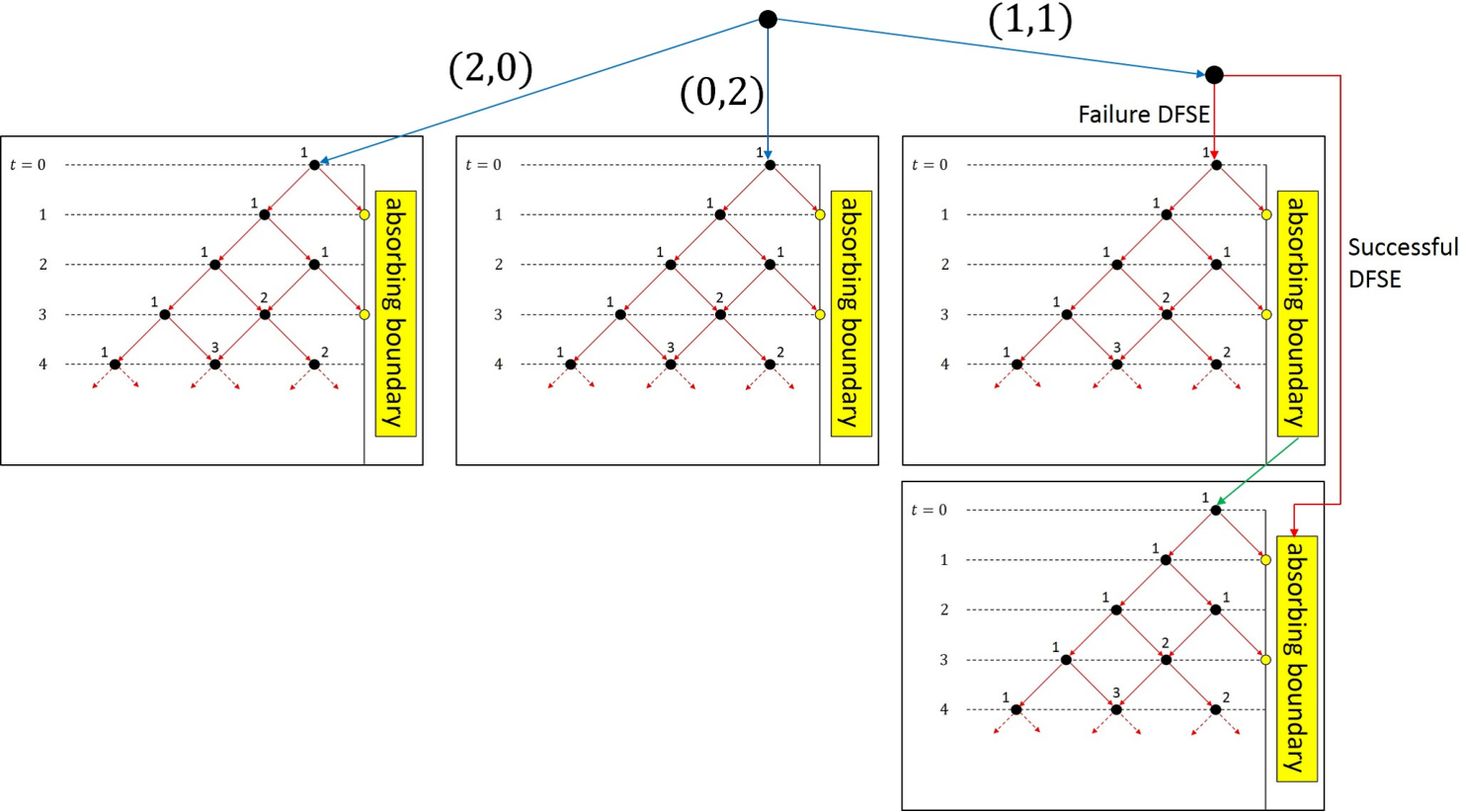}
\end{center}
\captionsetup{justification=raggedright,singlelinecheck=false}
\caption{The flow of the coherent-light-assisted DFS-BQC protocol. The pairs of numbers $(2,0)$, $(0,2)$, and $(1,1)$ represent the outcome of the first QND measurement in step (3)  $(N_s^{(2i-1,2i)},N_l^{(2i-1,2i)})$. In all three cases, when the walker arrives at the final absorbing boundary, $|+_{\theta_i}\rangle$ is prepared on Bob's side.}
\label{diagram}
\end{figure}

We calculate the success probabilities for each case (i)--(iii). The flow of the coherent-light-assisted DFS-BQC protocol is shown in Fig.~\ref{diagram}. We define $\mathcal{P}_n(T)\equiv e^{-\mu T}{(\mu T)^n}/{n!}$ and $\mathcal{T}_s\equiv{(|a|^2+|d|^2)}/{2}$. Here, $\mu$ and $T$ indicate the mean photon number of the coherent light sent by Alice to Bob, and the transmission rate of the quantum channel between Alice and Bob, respectively.

\begin{figure}[t]
 \begin{center}
  \includegraphics[width=8.6cm, clip]{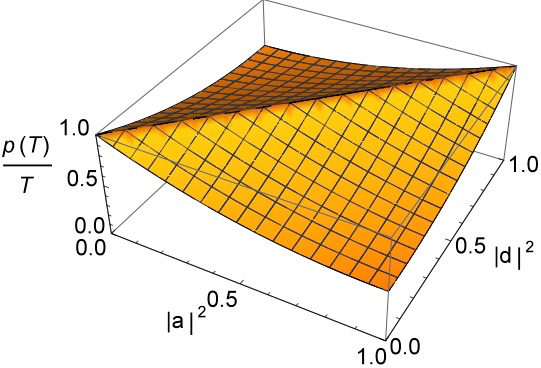}
\end{center}  
\captionsetup{justification=raggedright,singlelinecheck=false}
\caption{The coefficient $p(T)/T$ as a function of $|a|$ and $|d|$ 
in the large-$\mu$ limit.
}
\label{coefficient}
\end{figure}

\begin{enumerate}[(i)]
\item
Since each of the signal photon and single photons extracted from the coherent light is detected in mode $s$ with probability $\mathcal{T}_s$, 
the success probability of the repetition of the DFSEs in mode $s$ is
written by 
\begin{eqnarray*}
&&T\sum_{n=1}^{\infty}\mathcal{P}_n(T)\sum_{t=1}^{n}\binom{n-1}{t-1}\mathcal{T}_s^{t+1}(1-\mathcal{T}_s)^{n-t}\sum_{t'=0}^{\lfloor\frac{t-1}{2}\rfloor}C_{t'}q_0^{(\rm{i})}{Q_1}^{t'}\\
&&\equiv p_1(T).
\end{eqnarray*}
\item
The success probability of the repetition of the DFSEs in mode $l$ is written by
\begin{eqnarray*}
&&T\sum_{n=1}^{\infty}\mathcal{P}_n(T)\sum_{t=1}^{n}\binom{n-1}{t-1}(1-\mathcal{T}_s)^{t+1}(\mathcal{T}_s)^{n-t}\sum_{t'=0}^{\lfloor\frac{t-1}{2}\rfloor}C_{t'}q_0^{(\rm{ii})}{Q_2}^{t'}\\
&&\equiv p_2(T).
\end{eqnarray*}
\item
The success probability of the repetition of the DFSEs in modes $s$ and $l$ is written by
\begin{widetext}
\begin{eqnarray*}
&&2T\sum_{n=1}^{\infty}\mathcal{P}_n(T)\mathcal{T}_s(1-\mathcal{T}_s)q_0^{\rm{(iii-i)}}\\
&&+2T\sum_{n=3}^{\infty}\mathcal{P}_n(T)\mathcal{T}_s(1-\mathcal{T}_s)\sum_{t=1}^{n-2}\binom{n-1}{t}\mathcal{T}_s^t(1-\mathcal{T}_s)^{n-t-1}\Biggl\{\sum_{t'=0}^{\lfloor\frac{t-1}{2}\rfloor}C_{t'}{Q_1}^{t'+1}\Biggl\}\Biggl\{\sum_{t''=0}^{\lfloor\frac{n-t}{2}-1\rfloor}C_{t''}q_0^{\rm{(iii-ii)}}{Q_2}^{t''}\Biggl\}\\
&=&2T\sum_{n=1}^{\infty}\mathcal{P}_n(T)\mathcal{T}_s(1-\mathcal{T}_s)q_0^{\rm{(iii-i)}}\\
&&+2T\sum_{n=3}^{\infty}\mathcal{P}_n(T)\sum_{t=1}^{n-2}\binom{n-1}{t}\mathcal{T}_s^{t+1}(1-\mathcal{T}_s)^{n-t}\Biggl\{\sum_{t'=0}^{\lfloor\frac{t-1}{2}\rfloor}C_{t'}{Q_1}^{t'+1}\Biggl\}\Biggl\{\sum_{t''=0}^{\lfloor\frac{n-t}{2}-1\rfloor}C_{t''}q_0^{\rm{(iii-ii)}}{Q_2}^{t''}\Biggl\}\\
&\equiv&p_3(T).
\end{eqnarray*}
\end{widetext}
\end{enumerate}
The total success probability 
$p(T)$
of this protocol is
given as a function of the transmission rate $T$ 
by
\begin{eqnarray*}
p(T)=p_1(T)+p_2(T)+p_3(T).
\end{eqnarray*}
In the limit of $\mu\rightarrow\infty$,
we obtain
\begin{eqnarray*}
p(T)=&&T\Biggl\{\mathcal{T}_s^2\sum_{t'=0}^{\infty}C_{t'}q_0^{(\rm{i})}{Q_1}^{t'}+(1-\mathcal{T}_s)^2\sum_{t'=0}^{\infty}C_{t'}q_0^{(\rm{ii})}{Q_2}^{t'}\\
&&+2\mathcal{T}_s(1-\mathcal{T}_s)q_0^{(\rm{iii-i})}\\
&&+2\mathcal{T}_s(1-\mathcal{T}_s)\Biggl(\sum_{t'=0}^{\infty}C_{t'}{Q_1}^{t'+1}\Biggl)\Biggl(\sum_{t''=0}^{\infty}C_{t''}q_0^{(\rm{ii})}{Q_2}^{t''}\Biggl)\Biggl\}.
\end{eqnarray*}
The coefficient of $T$ is
\begin{widetext}
\begin{eqnarray*}
\frac{p(T)}{T}=&&\cfrac{1}{4}\Biggl\{2(|a|^2+|d|^2-2|ad|^2)-(2-|a|^2-|d|^2)^2\Bigl(\cfrac{||a|^2-|d|^2|}{2-|a|^2-|d|^2}-1\Bigl)-(|a|^2+|d|^2)^2\Bigl(\cfrac{||a|^2-|d|^2|}{|a|^2+|d|^2}-1\Bigl)\\
&&+(2-|a|^2-|d|^2)(|a|^2+|d|^2)\Bigl(\cfrac{||a|^2-|d|^2|}{2-|a|^2-|d|^2}-1\Bigl)\Bigl(\cfrac{||a|^2-|d|^2|}{|a|^2+|d|^2}-1\Bigl)\Biggl\},
\end{eqnarray*}
\end{widetext}
which is independent of 
$T$ in the large-$\mu$ limit and only depends on 
$|a|$ and $|d|$. 
Even when $\mu$ is finite, it is satisfied that the $T$ dependence of $p(T)$ is $O(T)$.
In Fig.~\ref{coefficient},
the coefficient $p(T)/T$ is plotted 
as a function of $|a|$ and $|d|$. Only when $|a|=|d|$ is satisfied, this coefficient becomes $1$ as the maximum. On the other hand, when $(|a|,|d|)=(1,0)$, or $(0,1)$, this coefficient becomes $1/2$ as the minimum.

\end{document}